# Far field photoluminescence imaging of single AlGaN nanowire in the sub-diffraction length scale using optical confinement of polarized light

A. K. Sivadasan,*,# Manas Sardar,$ and Sandip Dhara*,#

#Nanomaterials and Sensors Section, Surface and Nanoscience Division, Indira Gandhi Centre for Atomic Research, Kalpakkam-60310, India

$ Theoretical Studies Section, Materials Physics Division, Indira Gandhi Centre for Atomic Research, Kalpakkam-60310, India

*Abstract*

Till now the nanoscale focussing and imaging in the sub-diffraction limit is achieved mainly with the help of plasmonic field enhancement assisted with noble metal nanoparticles. Using far field imaging technique, we have recorded polarized spectroscopic photoluminescence (PL) imaging of a single AlGaN nanowire (NW) of diameter ~ 100 nm using confinement of polarized light. The nanowires on the substrate have a nematic ordering. It is found that the PL from a single NW is influenced by the proximity to other NWs with the PL intensity scaling as $1/(l \times d)$, where $l$ and $d$ are the NW length and the separation from the neighbouring NW, respectively. We show that this proximity induced PL intensity enhancement can be understood, if we assume the existence of reasonably long lived photons in the intervening space between the NWs. A nonzero non-equilibrium population of such photons causes stimulated emission leading to the enhanced PL emission with the intensity scaling as $1/(l \times d)$. The effect is analogous to the Purcell enhancement of polarized optical emissions induced by confined photons in micro-cavities. The enhancement of PL emission facilitated the far field spectroscopic imaging of a single semiconducting nanowire in the sub-diffraction regime.

---

Email : sivankondazhy@gmail.com; dhara@igcar.gov.in

Imaging a nanostructure or focusing to a dimension in the sub-diffraction limit, using the diffraction limited optical technique is a major achievement in recent times. The nanoscale imaging in the sub-diffraction limit was boosted with the application of plasmonics.[1,2] Surface plasmon resonance (SPR), originating from the coherent oscillation of conduction electrons of noble metal nanoclusters with the excitation of visible light, facilitate propagation of surface plasmon polaritons (SPP)[3,4] along with evanescent waves near the metal-dielectric interface to initiate imaging or focusing at sub-diffraction length scale.[5-11] Recently, polarized Raman studies of semiconducting nanostructures, using optically confined polarized light in dielectric contrast with surrounding media, has raised considerable interests in the scientific community.[12-15] The rate of interaction of a light emitting material with electromagnetic radiation can be enhanced by keeping it in a micro-cavity. The spontaneous emission rate is not only a material characteristic but also depends on the photon density (energy density of the fluctuating electromagnetic field) in the surrounding environment. The long lived cavity mode photons, enhances the optical density leading to the improved stimulated emission/absorption of light in the process of Purcell enhancement.[16-22] Tightly confined light or trapped photons in between two semiconducting or dielectric media with sizes comparable to the wavelength of the confined light, increases photon density in the micro-cavity leading to the enhanced light–matter interaction, enabling a range of nonlinear applications.[12-14,23-29] In the most recent study, polarized resonance Raman spectroscopy is successfully demonstrated to understand the crystallographic orientations of a single AlGaN nanowire (NW) in the sub-diffraction length scale.[15]

We report here, far field photoluminescence (PL) imaging of a single AlGaN NW, in the sub-diffraction length scale. We find the PL intensity coming from a single NW, is strongly influenced by the proximity of other NWs, even though the separation between the NWs is more than the exciting laser beam diameter. Increase in the PL intensity by an order from a single NW can be easily achieved by this proximity effect. In case of the optical confinement effects, a nonzero non-equilibrium population of long lived photons is envisaged to influence the increased PL emission intensity from the single NW.

AlGaN NWs with narrow size distribution with nematic ordering were synthesized using VLS process by an atmospheric pressure chemical vapor deposition technique with Au NP as catalyst.[16] Morphological features and crystallographic structural studies of the NWs were analyzed using a field emission scanning electron microscopy (FESEM; SUPRA 55 Zeiss) and a high resolution transmission electron microscopy (HRTEM; LIBRA 200FE Zeiss), respectively. The polarized PL studies were carried out using an excitation wavelength ($\lambda$) of 325 nm of He-Cd laser at room temperature (300K) with 2400 gr.mm$^{-1}$ grating for monochromatization. The spectra were collected using a near ultraviolet micro-spot focusing objective lens of magnification 40× having N.A. value of 0.50 (working distance of 1 mm) and assisted by a thermoelectrically cooled CCD detector in the backscattering configuration (inVia, Renishaw, UK). The resolved spot size of excitation laser beam ($D = 1.22\lambda/N.A.$) is expected to be ~ 0.8 μm. Each spectrum was recorded with laser power of 2.5 mW and with a 30 s integration time. The fully automated motorized sample stage, having a spatial resolution of 100 nm, was used for the polarized PL imaging purpose. The PL imaging was performed by integrating intensities, which was essentially the peak intensity distribution corresponding to a particular energy collected over a pre–defined area and grid resolution. In the present study, a total area of 3 x 1 μm$^2$ with 100 nm grid resolution was probed for the polarized micro–PL imaging.

The NWs with nematic ordering and a narrow size distribution of 100 (±10) nm are shown in the FESEM image [Fig. 1(a)]. The high resolution image [outset of Fig. 1(a)] of the cylindrically shaped AlGaN NWs shows very smooth surface morphology with Au catalyst nanoparticle (NP) at the tip of the NW [Fig. 1(a)]. The Au NPs participated in the vapor-liquid-solid (VLS)[16] growth process of the AlGaN NWs, which have uniform size of ~120 nm and was found to be well separated from each other. The interface region between the Au NPs and AlGaN NWs is negligible compared to the total area of the NW. The HRTEM image [Fig. 1(b)] is analyzed to show the interplanar spacing of 2.98 Å corresponding to *m*-plane (1-100) of the wurtzite phase indicating the growth direction of the single crystalline AlGaN NW along *m*–plane [1-100]. The selected area electron diffraction (SAED) pattern [Fig. 1(c)] is indexed to the wurtzite phase of crystalline AlGaN with zone axes along [0001].

The polarized PL measurements at 300K for the single AlGaN NW (Fig. 2) were performed in the two different configurations of $Z(XY)\bar{Z}$ and $Z(XX)\bar{Z}$ which are considered as perpendicular and parallel polarizations, respectively. In the former geometry, the electric field vector ($E_e$) is perpendicular to the horizontal *X*–axis ($E_e \perp X$, $\sigma$–polarization) and $E_e$ is parallel to the axis ($E_e \| X$, $\pi$–polarization) in the later configuration.[14,31,32] However, for both the cases, the incident direction (wave vector) of laser light is perpendicular to the horizontal axis ($k \perp X$) of AlGaN NW. It was found that polarized PL peak intensity value for a single NW is higher for $\pi$–polarization, as compared to that for the $\sigma$–polarization [Fig. 2(a)].[12-14,33] This observation is typical in NWs and is because of optical confinement of polarized light within the material, due to variation of refractive index of NWs and its surrounding.[12-14,18,24,25,30,33,34] The optical image of the monodispersed AlGaN NW with a separation of ~2 μm, using an objective of 100× with numerical aperture (N.A.) value of 0.8, is shown in the inset of Fig. 2(a). The single NW used for recording the PL spectra is indicated by an arrow. It was ensured that the individual NWs were well separated (~ 2 µm) compared to the excitation laser spot size of ~ 0.8 µm to receive a PL spectrum of a single NW alone. Considering around 5% error for the Gaussian spreading of incident beam diameter, the maximum possible laser beam diameter can be calculated as ~ 0.84 µm. Thus, it was ensured that the PL spectra was collected from a single NW with minimum interference of any other signals comes from nearest NWs. The peak observed at 3.55 eV is due to the free exciton (FE) recombination.[36,37] A tiny peak centered at 3.73 eV is assigned to 2-$A_1$(LO) mode of GaN.[36,37] The peak observed at 3.3 eV is identified with electronic transition from a shallow donor state of nitrogen vacancy ($V_N$) to a deep acceptor state of Ga vacancy ($V_{Ga}$)[38]. For the measured value of the FE at 3.55 eV, the Al content of NW was calculated ~ 2.93 at% by considering the room temperature band gap of GaN and AlN as 3.47 and 6.2 eV, respectively, in the band bowing formalism. The variation in intensity for different polarization configurations depends on the material geometry as well as the anisotropy of the medium. Thus the intensity of $\pi$–polarization ($I_\pi$) can be very different from that of the intensity of $\sigma$–polarization ($I_\sigma$) due the confinement effect as mention earlier. The observed polarization anisotropy is typically defined in terms of the polarization ratio ($\rho = I_\pi - I_\sigma / I_\pi + I_\sigma$; also called as the degree of polarization).[12-14,28,33,39] For the present study, the value of polarization ratio was calculated to be 0.533.

We observe a strong correlation of the PL intensity [Figs. 2(a) and 2(b)] with the proximity of other NWs. An increase in the PL intensity by one order is observed, when the separation from the nearest neighbor NW is decreased from 4 µm [inset Fig. 2(b)] to 2 µm [inset Fig. 2(a)]. We carried out the similar experiments for large number of times with different separation between the consecutive NWs to get a good statistics. We have plotted PL emission intensity at the peak energy versus the separation (*d*) between the NWs (Fig. 3). The inset of Figure 3 shows the typical PL spectra observed for few typical inter-NW (*d*) separations. It is very clear that the observed intensities of spectra are found to increase with decreasing *d* values. Interestingly the PL intensity varies inversely with the scaled mode area of (*l*×*d*), where *l* and *d* are the NW length and the inter-NW separation, respectively (Fig. 3). We have also confirmed $1/d^2$ dependence of the PL intensities with inter-NW separation for small cavities extending down to sub-micron size using scanning probe microscopy coupled to Raman spectrometer (supporting information Fig. S1).[40]

The nematic ordering of the AlGaN NWs with uniform size distribution [Fig. 1(a)] leads to the existence of long lived photons (owing to multiple reflections) in the intervening space between the NWs. These photons originate from external laser source as well as electronic transitions of PL within the materials, and exist so long as the external pump laser is on. The nonzero, non-equilibrium population density of such photon modes increase stimulated emission by utilizing electrons in exited states. The elevation in optical density of photons may be the reason for enhanced PL due to proximity of other NWs. We derived (supplementary information Quantitative Investigation of Proximity Effect Using Rate Equation)[40] an expression for the intensity of PL starting from the rate equation by considering the proximity effect between the NWs as

$$I_{PL} \propto R_P \tau_{sp}(1+2\sigma v_g N \tau_{ph}/V). \quad\quad\quad\quad\quad\quad\quad\quad\quad\quad\quad\quad\quad\quad(1)$$

where $R_P$ is the rate at which electrons are pumped to the upper level by absorption of photons. $\tau_{sp}$ is the time for spontaneous electronic transition from upper (*U*) to ground (*G*) state to emit a photon. The $N_U$ and $N_G$ be the number of electrons in the excited state and in the ground state ($N_G+N_U = N$, being a conserved quantity). The photon group velocity in the medium is $v_g$. The second term in the Eqn. (1) is responsible for the increase in PL intensity due to the presence of the intervening long lived photons in between the NWs. The increase in the PL intensity is proportional to lifetime of these photons $\tau_{ph}$, the stimulated emission cross section $\sigma$, and is inversely proportional to the volume ($V$) occupied by these photons. Since, $V = h \times l \times d$ where, $h$, $l$ and $d$ are the height, length and separation between the NWs. So, for a fixed $h$ (~100 nm), the diameter of the NWs lesser than the excitation wavelength ($\lambda$), the increase in PL intensity is inversely proportional to $l \times d$.

It can also be understood in a simplistic manner by considering the Purcell factor or figure of merit of the cavity resonator ($F_P \propto (\lambda/n)^3 Q/V_m$; in 3D) as the governing factor for the increase of emission which is directly proportional to the cube of excitation wavelength ($\lambda/n$) times the quality factor ($Q$; directly proportional to the confinement life time of photons) of the resonator. The $F_P$ is inversely proportional to the mode volume ($V_m$) determined essentially by the dimensions of the cavity and large value of $Q/V_m$ plays key role in the Purcell enhancement for light-matter interactions.[17-23] Even though the $Q$-factor for the open cavities is expected to be lower (absence of sharp PL peaks in Fig. 2 and inset in Fig. 3, as indicative) compared that for the closed ones, we observe a trend of PL intensity variation with respect to the mode area between the two NWs (Fig. 3). As a consequence of Purcell effect, dielectric micro-cavities are certainly notable for the observation and manipulation of strong light–matter coupling to semiconductors.[23] Considering uniform diameter of the NWs (fixed $h$ of the cavity) in our case, the optical density for the excitation photons is higher for the lower 2D cavity volumes. This may lead to the increased amount of transition rate leading to the Purcell enhancement in the PL emission for the lower 2D cavity volumes.[17,19,20]

The PL imaging of single AlGaN NW with an average diameter 100 (±10) nm is carried out over an area of 3 x 1 µm$^2$ using an excitation laser of wavelength of 325 nm. Figure 4 shows the PL maps of single NW generated by integrating the intensity of FE emission for $\pi$–polarization. The schematic of the experimental configuration is detailed in the supporting information Fig. S2.[40] The optical image of a single NW is also shown (inset Figure 4) with a schematic grid pattern used for the mapping. The scale bar of 3 µm represents the true dimension along the long axis of the NW only. The PL maps for $\sigma$–polarization with the variation of intensity along the NW for FE emission is also studied (supporting information Fig. S3).[40] In the Abbe's diffraction (sub-diffraction) limit ($\lambda$/2N.A.), the far field spectroscopic imaging is not possible for nanostructures having dimension <325 nm while using 325 nm excitation wavelength and an objective lens with N.A. value of 0.5. Considering negligible interface area of the catalyst Au NP, as compared to that for the total surface area of the NW [Fig. 1(a)], possibility of plasmonic interaction between the Au NPs and AlGaN NWs with a 325 nm excitation source can also be ruled out. However, the optical confinement effect for the light-matter interactions allow one to record the PL maps of single AlGaN NW in the sub-diffraction regime even in the absence of any plasmonic effect. The dimension of PL maps of single NW (~ 200 nm) using $\pi$–polarization is found to be enlarged as compared to the FESEM images of a single AlGaN NW [~100 nm; outset Fig. 1(a)]. The spread is essentially due to the focusing limit (~ 0.8 µm) of the incident beam which is higher than the spatial resolution (~100 nm) of the motorized sample stage used for the PL mapping.

In conclusion, the proximity induced enhancement of photoluminescence (PL) intensity is understood by assuming the existence of long lived photons in the intervening space between the nanowires (NWs), where a nonzero non-equilibrium population of such photons causes stimulated emission leading to enhanced PL emission with the intensity scaling as $1/(l \times d)$. The enhancement of electric field strength as well as the light-matter interactions can also be simultaneously understood as an effect of optical confinement in 2D micro-cavity due to the Purcell enhancement in the of 1D semiconductor NW-air-1D semiconductor NW columns. With the help of the enhanced PL emission, we record the far field spectroscopic imaging using the $\pi$–polarized PL spectra for a cylindrically shaped single AlGaN NW with a diameter 100 ($\pm$10) nm in the sub-diffraction regime without the aid of any plasmonic effects. Thus, the report is encouraging for analysing and processing nanoscopic objects in the sub-diffraction limit with the far field configuration using the diffraction limited optics. The proposed alternative methodology avoids both the complex material processing for achieving optimized noble metal thickness involved in the plasmonic technique.

One of us (AKS) acknowledges the Department of Atomic Energy for allowing him to continue the research work. We thank S. Amirthapandian, and S. Polaki, IGCAR for their help in TEM and FESEM studies, respectively.

**References**

1. A. Polman, Science **322**, 868 (2008).

2. E. Ozbay, Science **311**, 189 (2006).

3. S. Dhara, C.-Y. Lu, P. Magudapathy, Y.-F. Huang, W.-S. Tu, K.-H. Chen, Appl. Phys. Lett. **106**, 023101 (2015).

4. S. Dhara, C.-Y. Lu, K.-H. Chen, Plasmonics **10**, 347 (2015).

5. S. A. Maier, P. G. Kik, H. A. Atwater, S. Meltzer, E. Harel, B. E. Koel, A. A. G. Requicha, *Nat. Mater.* **2**, 229 (2003).

6. P. Andrew, W. L. Barnes, Science **306**, 1002 (2004).

7. S. I. Bozhevolnyi, V. S. Volkov, E. Devaux, J.-Y. Laluet, T. W. Ebbesen, Nature **440**, 508 (2006).

8. H. J. Lezec, A. Degiron, E. Devaux, R. Linke, L. Martin-Moreno, F. Garcia-Vidal, T. W. Ebbesen, Science **297**, 820 (2002).

9. J. B. Pendry, Phys. Rev. Lett. **85**, 3966 (2000).

10. N. Fang, H. Lee, C. Sun, X. Zhang, Science **308**, 534 (2005).

11. S. W. Hell, Science **316**, 1153 (2007).

12. H. E. Ruda, A. Shik, Phys. Rev. B **72**, 115308 (2005).

13. H. E. Ruda, A. Shik, J. Appl. Phys. **100**, 024314 (2006).

14. J. Wang, M. S. Gudiksen, X. Duan, Y. Cui, C. M. Lieber, Science **293**, 1455 (2001).

15. A. K. Sivadasan, A. Patsha, S. Dhara, Appl. Phys. Lett. **106,** 173017 (2015).

16. A. K. Sivadasan, A. Patsha, S. Polaki, S. Amirthapandian, S. Dhara, A. Bhattacharya, B. K. Panigrahi, A. K. Tyagi, Crystal Growth & Design **15**, 1311 (2015)

17. M. Boroditsky, R. Vrijen, T. F. Krauss, R. Coccioli, R. Bhat, E. Yablonovitch, J. Lightwave Technol. **17**, 2096 (1999).

18. L. Sapienza, H. Thyrrestrup, S. Stobbe, P. D. Garcia, S. Smolka, P. Lodahl, Science **327**, 1352 (2010).

19. K. J. Vahala, Nature **424**, 839 (2003).

20. J.-M. Gerard, B. Gayral, J. Lightwave Technol. **17**, 2089 (1999).

21. S. J. Yoo, Q.-H. Park, Phys. Rev. Lett. **114**, 203003 (2015).

22. J. Li, D. Fattal, M. Fiorentino, R. G. Beausoleil, Phys. Rev. Lett. **106**, 193901 (2011).

23. B. Piccione, C. O. Aspetti, C.-H. Cho, R. Agarwal, Rep. Prog. Phys. **77**, 086401 (2014).

24. T. Xu, S. Yang, S. V. Nair, H. E. Ruda, Phys. Rev. B **75**, 125104 (2007).

25. J. S. Foresi, P. R. Villeneuve, J. Ferrera, E. R. Thoen, G. Steinmeyer, S. Fan, J. D. Joannopoulos, L. C. Kimerling, H. I. Smith, E. P. Ippen, Nature **390**, 143 (1997).

26. G. S. Wiederhecker, C. M. B. Cordeiro, F. Couny, F. Benabid, S. A. Maier, J. C. Knight, C. H. B. Cruz, H. L. Fragnito, Nat. Photonics **1**, 115 (2007).

27. T. Tanabe, M. Notomi, E. Kuramochi, A. Shinya, H. Taniyama, Nat. Photonics **1**, 49 (2006).

28. D. van Dam, D. R. Abujetas, R. Paniagua-Dominguez, J. A. Sanchez-Gil, E. P. A. M. Bakkers, J. E. M. Haverkort, J. Gómez-Rivas, Nano Lett. **15**, 4557 (2015).

29. T. Khudiyev, E. Ozgur, M. Yaman, M. Bayindir, Nano Lett. **11**, 4661 (2011).

30. R. F. Cregan, B. J. Mangan, J. C. Knight, T. A. Birks, P. S. Russell, P. J. Roberts, D. C. Allan, Science **285**, 1537 (1999).

31. P. P. Paskov, T. Paskova, P.-O. Holtz, B. Monemar, Phys. Rev. B **70**, 035210 (2004).

32. J. B. Schlager, N. A. Sanford, K. A. Bertness, J. M. Barker, A. Roshko, P. T. Blanchard, Appl. Phys. Lett. **88**, 213106 (2006).

33. H.-Y. Chen, Y.-C. Yang, H.-W. Lin, S.-C. Chang, S. Gwo, Opt. Express **16**, 13465 (2008).

34. Y. Bian, Q. Gong, Nanotechnology **25**, 345201 (2014).

35. B. Wood, J. B. Pendry, D. P. Tsai, Phys. Rev. B **74**, 115116 (2006).

36. K. K. Madapu, S. Dhara, S. Amirthapandian, S. Polaki, A. K. Tyagi, J. Phys. Chem. C **117**, 21930 (2013).

37. A. Patsha, P. Sahoo, S. Dhara, S. Amirthapandian, A. K.Tyagi, J. Raman Spectrosc. **44**, 651 (2013).

38. M. A. Reshchikov, H. Morkoç, J. Appl. Phys. **97**, 061301 (2005).

39. L. Rigutti, M. Tchernycheva, A. D. L Bugallo, G. Jacopin, F. H. Julien, F. Furtmayr, M. Stutzmann, M. Eickhoff, R. Songmuang, F. Fortuna, Phys. Rev. B **81**, 045411 (2010).

40. See supplementary material at [URL will be inserted by AIP] for additional information.

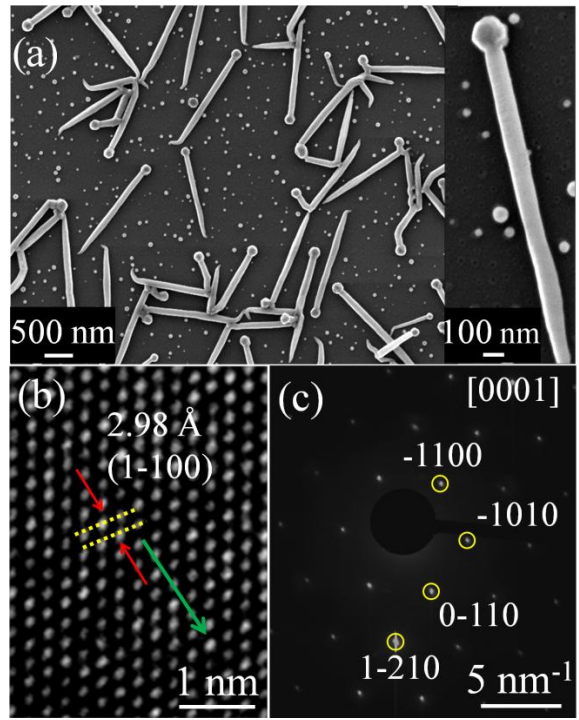

**FIG. 1.** (a) Monodispersed AlGaN NWs with a narrow size distribution of 100 (±10). Outset shows a typical cylindrically shaped single NW with smooth surface. (b) HRTEM image of a single AlGaN NW shows growth direction along *m*–plane [1-100]. (c) The SAED pattern is indexed to wurtzite phase of AlGaN with zone axis along [0001].

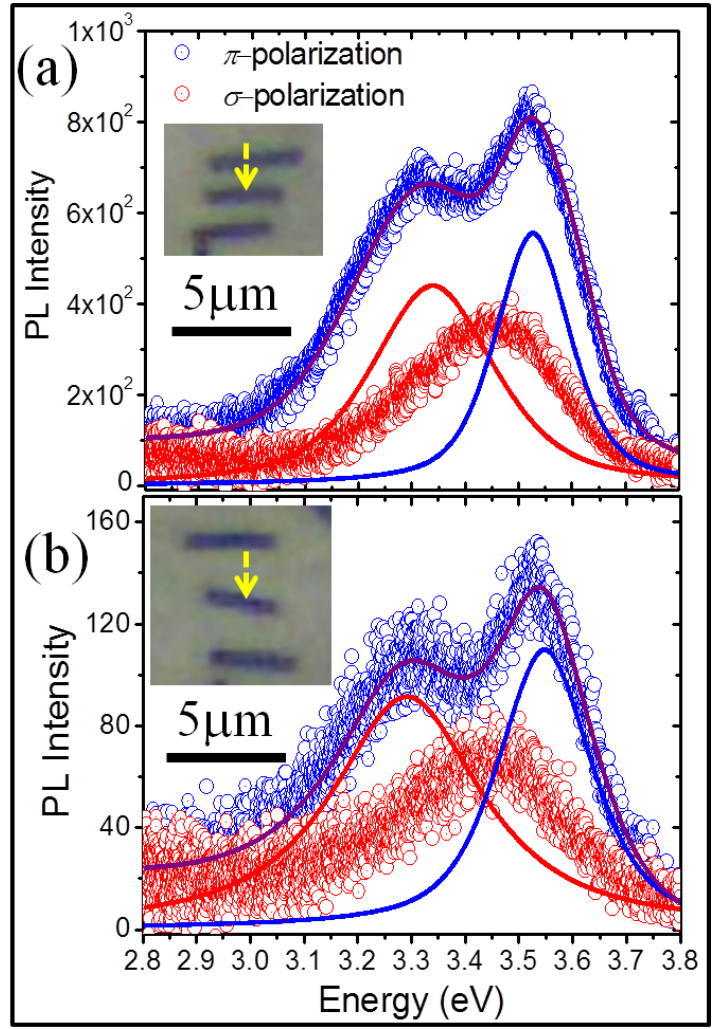

**FIG. 2.** Polarized PL spectra for π– and σ–polarized configurations of a single AlGaN NW for the inter-NW separations of (a) 2 μm and (b) 4 μm. Inset shows optical image of monodispersed NWs with different inter-NW separations on the Si (100) substrate. The single NW selected for the polarized PL spectra is indicated by an arrow. The scale bar corresponds to a length of 5 μm.

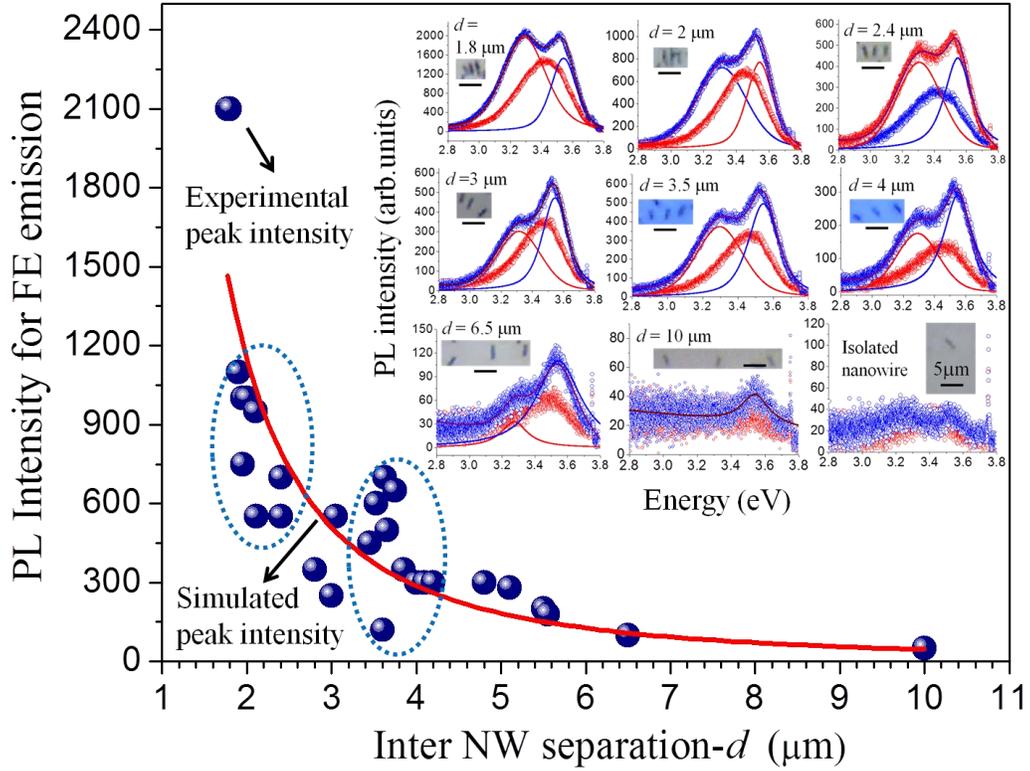

**FIG. 3.** The statistical dependence of PL intensity in the 2D semiconductor micro-cavities with respect to the separation of consecutive nanowires. Inset shows the photoluminescence spectra correspond to different separations and the scale bars in the inset correspond to a length of 5 μm.

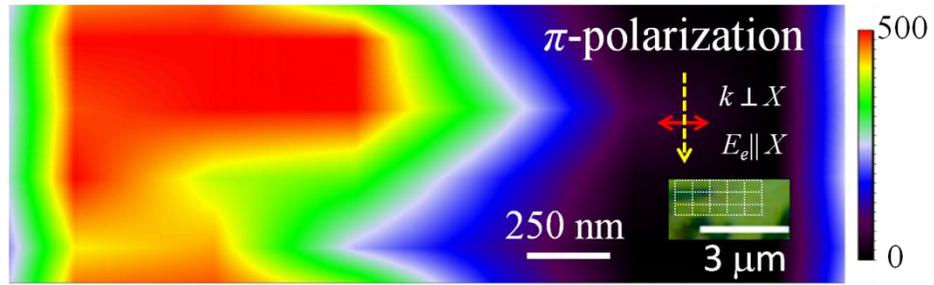

**FIG. 4.** The π-polarized PL maps for a single AlGaN NW. The optical image of a typical single NW with a schematic grid pattern is shown in the inset.

**Supporting Information**

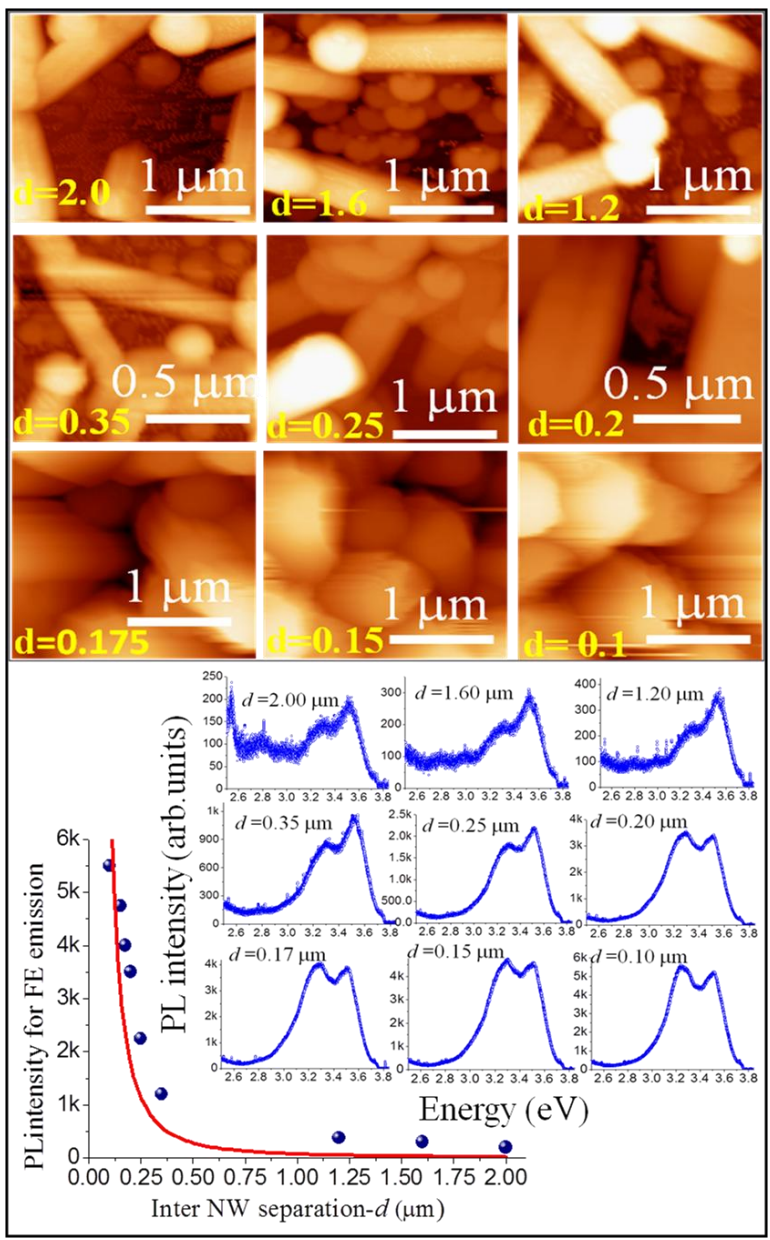

**Figure S1.** The statistical dependence of photoluminescence (PL) intensity in the 2D semiconductor micro- and nano-cavities with respect to the separation of consecutive nanowires. Inset shows the photoluminescence spectra correspond to different separations. The PL intensities are recorded using an atomic force microscopic tip of 20 nm assisted with a long

working distance near ultraviolet objective of 15× (N.A. = 0.4) in the scanning probe microscopy (SPM, MV4000; Nanonics, Israel) coupled to Raman spectrometer (inVia; Renishaw, UK).

**Quantitative Investigation of Proximity Effect Using Rate Equation**

In order to understand the phenomenon quantitatively, let us assume that the external pump laser promotes electrons to upper level (*U*) from the ground state (*G*), where $N_U$ and $N_G$ be the number of electrons in the excited state and in the ground state ($N_G+N_U = N$, being a conserved quantity). One can write down a rate equation for spontaneous emission/absorption of photons, $dN_U/dt = -dN_G/dt$,

$$\frac{dN_U}{dt} = -\frac{N_U}{\tau_{sp}} - W_P(N_U - N_G) + R_P \quad \ldots \ldots (1)$$

where, $\tau_{sp}$ is the time for spontaneous electronic transition from *U* to *G* state, with emission of a photon. $R_P$ is the rate at which electrons are pumped to the upper level by absorption of photons. The intermediate term is necessary when the electronic transitions are also induced by external photons, *i.e.*, stimulated emission/absorption processes. $W_P$ is the rate of stimulated emission/absorption per electron. Suppose, $\sigma$ is the cross sectional area for stimulated emission from a single electron at upper level due to the interaction with one external photon, and $N_L$ be the total number of local photons, confined in a volume *V*. If the photon group velocity in the medium is $v_g$, then the total number of stimulated emission/absorption events per second from one electron is $\sigma \cdot v_g \times N_L/V = W_P$ (having dimension inverse of time). When the system is in dynamic equilibrium, putting $\frac{dN_U}{dt} = 0$ one gets,

$$N_U = \frac{R_P + \frac{(\sigma.v_g)NN_L}{V}}{\frac{2(\sigma.v_g)N_L}{V} + \frac{1}{\tau_{sp}}} \quad \ldots \ldots (2)$$

If there are no local photons (no stimulated emission or absorption of photons), *i.e*, $N_L = 0$, the time averaged $N_U = R_P \tau_{sp}$. On the other hand, in large $N_L$ limit, $N_U$ approaches the value $N/2$, which is the maximum thermodynamically achievable population of the upper level. Similarly we can write down a rate equation for the number of photons responsible for stimulated emission/absorption process,

$$\frac{dN_L}{dt} = -\frac{N_L}{\tau_{ph}} + \frac{N_U}{\tau_{sp}} - \frac{(\sigma \cdot v_g)}{V}(N_U - N_G) \dots\dots\dots(3)$$

where, $\tau_{ph}$ is the decay time for these photons due to processes other than the stimulated emission/absorption. In order to get the steady state value of $N_L$, we set $dN_L/dt=0$. For $N_G = N - N_U$, and expressing $N_U$ in terms of $N_L$ from the earlier equation (2), we find $N_L = R_P \tau_{ph}$ and

$$N_U = R_P \tau_{sp} \frac{1 + \frac{2(\sigma \cdot v_g) N \tau_{ph}}{V}}{1 + \frac{(\sigma \cdot v_g) \tau_{ph} \cdot R_P \tau_{sp}}{V}} \dots\dots\dots(4)$$

The PL intensity, $I_{PL}$ is the total number of photons per second coming from all possible transitions of electrons from upper to lower level. The number of such transition per second is proportional to $N_U$, that is, $I_{PL} = (h\upsilon)|M|^2 N_U$, where $h\upsilon$ is the energy difference between $U$ and $G$, and $M$ being matrix element for this transition. The ratio of the second term in denominator and numerator, is $R_P \tau_{sp}/N = N_U(N_L=0)/N \ll 1$. In fact this ratio approaches 1/2 only when $N_L$ approaches infinity. Neglecting the second term in the denominator, we get,

$$I_{PL} \propto R_P \tau_{sp}(1+2\sigma v_g N \tau_{ph}/V). \dots\dots\dots(5)$$

The second term in the equation (5) is responsible for the increase in PL intensity due to the presence of the intervening long lived photons in between the NWs. The increase in the PL intensity is proportional to lifetime of these photons $\tau_{ph}$, the stimulated emission cross section $\sigma$, and is inversely proportional to the volume ($V$) occupied by these photons. Since, $V = h \times l \times d$ where, *h*, *l* and *d* are the height, length and separation between the NWs. So, for a fixed *h* (~100 nm, the diameter of the NWs lesser than the excitation wavelength λ), the increase in PL intensity is inversely proportional to *l*×*d*.

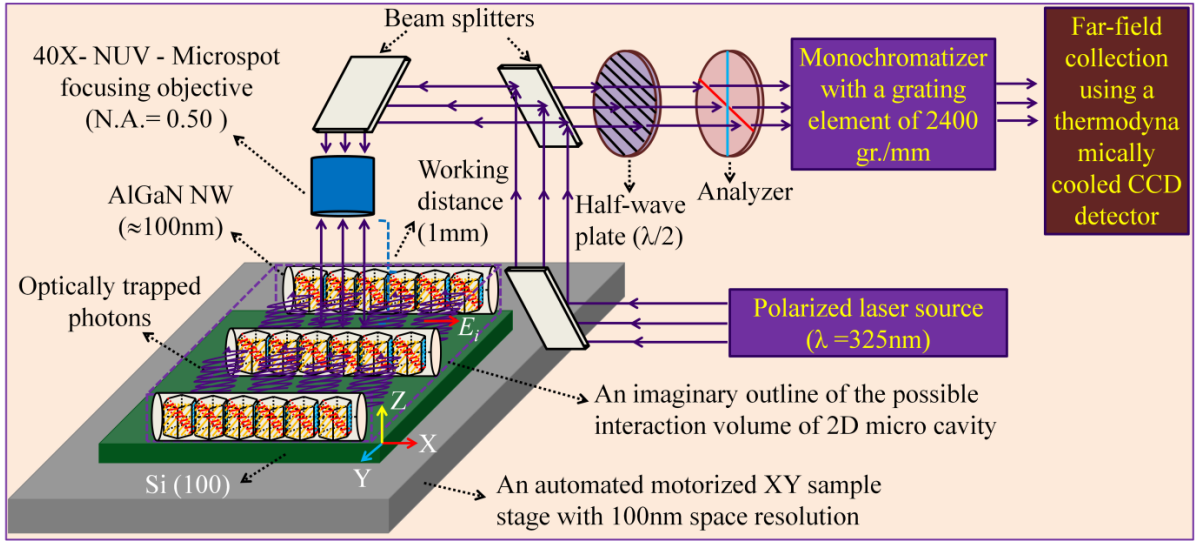

**Figure S2.** The schematic representation of experimental set up for taking single nanowire spectra and imaging with an automated polarized photoluminescence spectroscopy. The imaginary outline of possible 3D interaction volume depicts the trapped photons leading to the optical confinement effect in between the semiconducting AlGaN nanowire columns. The representative plane waves are bounce back and forth in between the arrays of nanowires inside the defined 3D interaction volume.

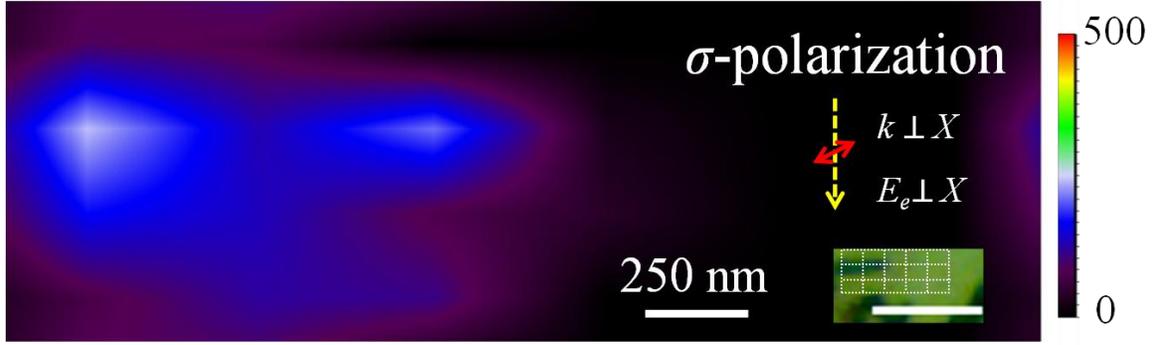

**Figure S3.** The photoluminescence mapped image for the intensity of FE emission collected from the defined area 3 x 1 μm² and the FE intensity distribution of single NW shown inside the rectangular dotted contour. The σ–polarized PL maps for a single AlGaN NW. The scale bar corresponds to a length of 3 μm.

The variation of intensity in the photoluminescence (PL) maps observed for different polarization configuration is due to difference in the interaction of electric field strength with NW. In π–polarization, the excitation electric field which is parallel to the NW long axis will completely interact with the NW (Fig. 4 in the manuscript). In this configuration, the internal electric field $E_{i\pi}$ in the NW medium remains same as the external electric field $E_{e\pi}$ (ie; $E_{i\pi} \sim E_{e\pi}$). Whereas in case of σ–polarization, the internal electric field $E_{i\sigma}$ will attenuate according to the relation $E_{i\sigma} = [2\varepsilon_0 / (\varepsilon + \varepsilon_0)]E_{e\sigma}$, where $E_{e\sigma}$ is the external electric field perpendicular to the long axis, $\varepsilon$ is the dielectric constant of the cylindrical NW (= 5.8 for GaN) and $\varepsilon_0 = 1$ in vacuum. Along with the cylindrical geometry of the NW, the large dielectric contrast between the NW and the surrounding medium strongly favors absorption and emission of light for π–polarization compared to those for σ–polarization.[1,2] The intensity distributions in the PL maps for FE emissions with different polarizations are observed to be good agreement with the above relation.

# References

1. J. Wang, M. S. Gudiksen, X. Duan, Y. Cui, and C. M. Lieber, Science **293**, 1455 (2001)

2. J. B. Schlager, N. A. Sanford, K. A. Bertness, J. M. Barker, A. Roshko and P. T. Blanchard, Appl. Phys. Lett. **88**, 213106 (2006).